\documentclass[%
 aip,amsmath,amssymb,
 reprint]{revtex4-1}

\usepackage{graphicx}
\usepackage{color}
\usepackage{dcolumn}
\usepackage{bm}
\usepackage[utf8]{inputenc}
\usepackage[T1]{fontenc}
\usepackage{etoolbox}
\usepackage{epstopdf}
\usepackage{textcomp}
\usepackage{natbib}
\usepackage[pagewise]{lineno}

\makeatletter

\newcommand{\Rmnum}[1]{\expandafter\@slowromancap\romannumeral #1@}
\makeatother
\begin{document}


\title{Impedance-matched high-overtone bulk acoustic resonator}
\author{M. Kurosu}
\email{megumi.kurosu@ntt.com}
\affiliation{NTT Basic Research Laboratories, NTT Corporation, Atsugi-shi, Kanagawa 243-0198, Japan}

\author{D. Hatanaka}
\affiliation{NTT Basic Research Laboratories, NTT Corporation, Atsugi-shi, Kanagawa 243-0198, Japan}

\author{R. Ohta}
\affiliation{NTT Basic Research Laboratories, NTT Corporation, Atsugi-shi, Kanagawa 243-0198, Japan}

\author{H. Yamaguchi}
\affiliation{NTT Basic Research Laboratories, NTT Corporation, Atsugi-shi, Kanagawa 243-0198, Japan}

\author{Y. Taniyasu}
\affiliation{NTT Basic Research Laboratories, NTT Corporation, Atsugi-shi, Kanagawa 243-0198, Japan}

\author{H. Okamoto}
\affiliation{NTT Basic Research Laboratories, NTT Corporation, Atsugi-shi, Kanagawa 243-0198, Japan}

\date{\today}

\begin{abstract}
We demonstrated a nearly impedance-matched high-overtone bulk acoustic resonator (HBAR) operating at super high frequency ranges, using an epitaxial AlN piezoelectric layer directly grown on a conductive SiC cavity substrate with no metal layer insertion.
The small impedance mismatch was verified from the variation in the free spectral range (FSR); the experimentally obtained FSR spectra was greatly reproduced by using the Mason model.
Broadband phonon cavity modes up to the K-band (26.5 GHz) were achieved
at an AlN layer thickness of 200 nm. 
The high figure of merit of $f\times\text{Q} = 1.3\times 10^{13}\ \textrm{Hz}$ at 10 GHz was also obtained.
Our nearly impedance-matched high-quality HBAR will enable the development of microwave signal processing devices for 5G and future 6G communication systems, such as low-phase noise oscillators and acoustic filters, as well as research on high-frequency acoustic systems hybridized with electric, optical, and magnetic systems.
\end{abstract}

\maketitle
\textcolor{black}{High-overtone bulk acoustic resonators (HBARs)} have scalability, high frequency, high quality (Q)-factor, and multiple phonon modes \cite{lakin1981acoustic,lakin1993high, Bailey1992frequency, machado2019generation, daugey2015high,wu2021new}. Therefore, they have been used for applications such as low-phase-noise frequency reference oscillators and material characterization \cite{Driscoll1992, Zhou6202419,Zhang2006,kongbrailatpam2020effects,rabus2015high}. HBARs have been investigated for quantum applications since multiple modes can interact with quantum sources \cite{gokhale2020epitaxial}, such as superconducting qubits \cite{chu2017quantum,kervinen2018interfacing} and color centers \cite{macquarrie2015coherent,chen2019engineering}, and the information can be stored in the HBARs.

A conventional HBAR consists of a piezoelectric thin layer sandwiched between the top and bottom metal electrodes on a thicker handle substrate.
In this structure, bulk acoustic waves (BAWs) are excited through the inverse-piezoelectric effect by applying an alternating voltage between the top and bottom electrodes.
An excited BAW longitudinally propagates and is reflected at the top and bottom interfaces of the HBAR, which results in multiple phonon modes.
\textcolor{black}{To excite the BAWs in the GHz range, piezoelectric materials with small dielectric and acoustic loss, such as AlN\cite{hashimoto2009rf}, AlScN\cite{Sorokin2021Toward,Sotnikov2021Microwave}, Ba$_{0.5}$Sr$_{0.5}$TiO$_3$\cite{kongbrailatpam2020effects}, and ZnO\cite{baumgartel2009experimental}, are commonly used.}

In conventional HBARs, however, the bottom metal electrodes inserted between the piezoelectric layer and acoustic cavity prevents smooth acoustic-impedance transition, which results in inefficient acoustic power transfer \cite{gokhale2020engineering}. 
The impedance mismatch causes frequency variation in the mode spacing, called free spectral range (FSR) variation \cite{zhang2003resonant}. 
An HBAR with large FSR variation is not convenient for applications such as information storage and broadband material spectroscopy.
To solve these problems with conventional HBARs, we fabricated a nearly impedance-matched HBAR with an 800-nm-thick epitaxial AlN layer directly grown on an n-type conductive \textcolor{black}{6H-SiC} (n-SiC) substrate by metal organic chemical vapor deposition (MOCVD). 
The bottom metal electrode is not necessary since the conductive substrate acts as both a floating bottom electrode and acoustic cavity. 
The nearly matched acoustic impedance between the AlN layer and n-SiC substrate was experimentally confirmed from the small variation in FSR. The normalized FSR variation factor in our 800-nm AlN HBAR was $8.7\times 10^{-4}$,  which is comparable to the best value in previously reported HBARs.
We conducted a numerical calculation using the one-dimensional Mason model and confirmed that the experimentally obtained FSR spectra were well described with this model. 
The excellent acoustic impedance matching between the AlN layer and n-SiC substrate without the bottom metal electrode resulted in low FSR variation.
Furthermore, high-frequency operation up to the K-band (26.5 GHz) was demonstrated by reducing the thickness of the AlN layer. The high figure of merit of $f\times\text{Q} = 1.3\times 10^{13}\ \textrm{Hz}$ at 10 GHz was also achieved.

A piezoelectric AlN (800 nm) layer was epitaxially grown on a double-side polished n-SiC substrate by MOCVD \cite{taniyasu2007threading}. The n-SiC substrate has a resistivity of 0.071 $\Omega\cdot \text{cm}$, thickness of 244 $\mu$m, and [0001] orientation. The epitaxial AlN layer is c-axis oriented and has Al-polarity. The full-width at half-maximum values of x-ray rocking curves were 0.143$^\circ$ for the AlN (0002) plane and 0.146$^\circ$ for the AlN (10-12) plane, respectively.
The structure of our HBAR is shown in Fig. 1(a), where the top metal electrode (Ti/Al = 5/100 nm) is deposited by electron beam evaporation after the growth of the AlN layer and used for the ground-signal-ground (GSG) probe. The area of the signal pad determines the active area of the HBAR. The active area is 900 $\mu$m$^2$ unless otherwise mentioned.
The n-SiC substrate plays both roles of a floating bottom electrode and low-loss acoustic cavity. 
A  scanning electron microscopy image of the cross-section of our HBAR is shown in the inset of Fig. 1(a), and the top view is shown in Fig. 1(b).

Acoustic impedance matching is vital for efficient acoustic power transfer from the piezoelectric layer to acoustic cavity \cite{gokhale2020engineering}. Since all acoustic energy is generated in the piezoelectric layer, it is important that the acoustic impedance of the material stack from the piezoelectric layer to substrate should be matched. When an acoustic wave is reflected at the interface between materials 1 and 2, the fractional reflected ($R$) and transmitted ($T$) power is expressed as
\begin{eqnarray} 
	R=\left| \frac{Z_1-Z_2}{Z_1+Z_2}\right|^2,\ \ T = 1-R,
\end{eqnarray}
where $Z_i = S_i\rho_i v_i$ denotes the acoustic impedance of a material, and normal transport of longitudinal acoustic wave is assumed. The notations $S_i$, $\rho_i$, and $v_i$ represent the area, density, and longitudinal velocity of a material.
The ratio of the acoustic power transmission from a piezoelectric layer (0$^\text{th}$ layer) to the $i^\text{th}$ layer is calculated as follows \cite{gokhale2020engineering}.
\begin{eqnarray} 
	P_i = P_0\times \prod_1^i T_{(i-1)\rightarrow i}
	\label{e1}
\end{eqnarray}
where $P_0$ denotes the generated acoustic power in the piezoelectric layer, and lossless materials are assumed.
Conventional HBARs have a metal/piezoelectric material/metal/cavity structure.
The ratio of the acoustic power transmission is calculated in conventional HBARs with the Al/Ti/AlN/M/n-SiC structure, as shown in Fig. 1(c), where M denotes commonly used metals such as Au, Al, and Mo.
The acoustic impedance of such metals differ from that of AlN and SiC. Due to the impedance mismatch, 71$-$89 \% of the acoustic power generated in the AlN layer reaches the SiC substrate.
We calculated this value in our HBAR with the Al/Ti/AlN/n-SiC structure, as shown in Fig. 1(d). The impedance matching between the AlN layer and n-SiC substrate leads to efficient acoustic power transfer (\textgreater 99 \%). 
\textcolor{black}{We also compared the lattice constant and fractional transmitted power for commonly used acoustic cavities such as sapphire, Si, quartz, and diamond (see Supplementary material). Of these materials SiC has the lattice constant closest to AlN and highest transmitted power.}
Since it is difficult to directly measure the acoustic power transmission in an HBAR, an experimental evaluation of acoustic impedance matching was conducted involving measuring the FSR variation.

\begin{figure}[htp]
	\centering
	\includegraphics[width=7cm]{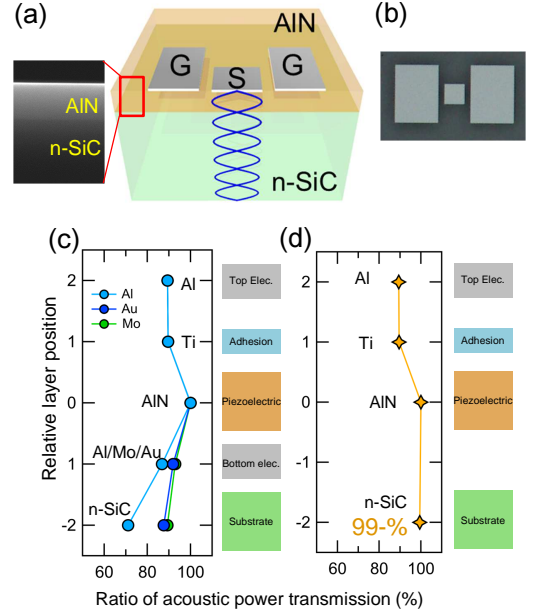}
	\caption{(a) Schematic showing our HBAR. Epitaxial AlN layer is directly grown on n-SiC substrate. Inset shows cross-sectional scanning electron microscopy image of HBAR.
	(b) Microscope photograph of top view of HBAR. (c)–(d) Ratios of acoustic power transmission. Vertical axis shows relative layer position with respect to piezoelectric layer. (c) Al/Ti/AlN/M/n-SiC, where M indicates metal (Al, Mo, Au), (d) Al/Ti/AlN/n-SiC (ours).}
	\label{f1}
\end{figure}  

\begin{figure*}[t]
	\centering
	\includegraphics[width=15cm]{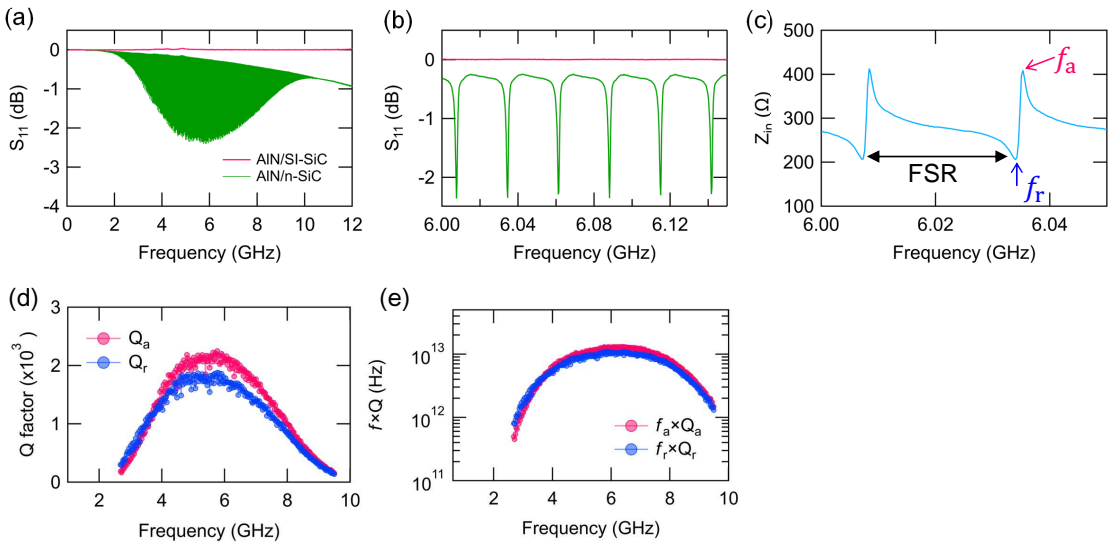}
	\caption{HBAR characterization. (a) Reflection spectrum of our HBAR with n-SiC and SI-SiC substrates. Multiple HBAR modes were measured with n-SiC substrate (green). In contrast, no HBAR mode was observed with SI-SiC substrate (red). Active area of HBARs was 900 $\mu$m$^2$. (b) Enlarged view of $S_{11}$. (c) Electric input impedance, which has local minimum (maximum) at resonance (anti-resonance). Black arrow indicates free spectral range. 
	(d) Frequency dependence of Q-factor. Each circle indicates individual HBAR mode. Blue (red) circle corresponds to resonance (anti-resonance). (e) Frequency dependence of $f\times\text{Q}$ product.}
	\label{f2}
\end{figure*}

We measured the reflection spectrum ($S_{11}$) of our HBAR by using the GSG probe and a vector network analyzer (VNA) (Keysight N5222B) at room temperature and ambient pressure. The maximum operating frequency of the VNA is up to 26.5 GHz. The measurement system was calibrated using the standard open-short-load method. We first examined if the n-SiC substrate works as a floating bottom electrode. For comparison, we fabricated an additional HBAR in which the substrate was changed to semi-insulating SiC (SI-SiC) and the piezoelectric AlN layer was epitaxially grown on it. The material stack of this other HBAR is Al/Ti/AlN/SI-SiC. 
Figure 2(a) shows the $S_{11}$ of both HBARs, where the green and red lines indicate the HBARs with the n-SiC and SI-SiC substrates, respectively. 
It is clear that a BAW was generated in the HBAR with the n-SiC substrate. In contrast, no signal was measured in the HBAR with the SI-SiC substrate. These results confirm that the n-SiC substrate effectively works as a floating bottom electrode.

The GHz acoustic vibration was generated through the inverse piezoelectric effect of AlN by applying the RF voltage between the signal pad and n-SiC substrate. Our HBAR exhibited multiple phonon modes as shown in Fig. 2(b), where the periodic frequency spacing between the modes are observed. The resonance frequency of the $m^\text{th}$ HBAR mode is given by,
\begin{eqnarray} 
	f_m \sim m\times \left( \frac{v_\textrm{s}}{2t_\textrm{s}} \right),
\end{eqnarray}
where $v_\textrm{s}$ and $t_\textrm{s}$ denote longitudinal acoustic velocity and thickness of the substrate, respectively.
The Q-factor of the HBAR is important for its applications and is deduced from the electric input impedance ($Z_\text{in}$), which is calculated from the reflection as follows.
\begin{eqnarray} 
	Z_\text{in} = \frac{1+S_{11}}{1-S_{11}}Z_0,
\end{eqnarray}
where $Z_0$ is the system impedance of a measurement path and is 50 $\Omega$. 
Each individual HBAR mode has a resonance and anti-resonances ($f_\text{r}$, $f_\text{a}$), where the $Z_\text{in}$ has a local minimum (maximum) at the resonance (anti-resonance) as shown in Fig. 2(c). The frequency spacing of each mode corresponds to the FSR. The Q-factor of these resonances is defined by
\begin{eqnarray} 
	\text{Q}_\text{ r, a} = \frac{f}{2}\left|\frac{\partial \angle Z_\text{in}}{\partial f}\right|_\text{r, a}.
\end{eqnarray}
Figure 2(d) shows the Q-factors of each HBAR mode, where the maximum Q-factor is about 2200 at around 5.8 GHz. 
\textcolor{black}{Our semiconductor's bottom electrode has lower conductance than those of metal electrodes, which may degrade the Q-factor due to Joule heating\cite{hashimoto2009rf}. A semiconductor substrate with high doping concentration may improve the Q-factor in the future.}

The product of the frequency and measured Q-factor ($f\times\text{Q}$ product) is a figure of merit, as shown in Fig. 2(e). 
\textcolor{black}{Although the $f\times\text{Q}$ product of $1.3\times 10^{13}$ Hz was achieved at 6 GHz, the calculated theoretical energy loss limit in the Landau-Rumer regime for n-SiC is $5.7\times 10^{14}$ Hz (see Supplementary material). This discrepancy indicates that the loss mechanism of our HBAR is not dominated by phonon-phonon dissipation \cite{tabrizian2009effect} but includes other dissipation such as ohmic loss in the electrode, leaking waves as the transverse mode, and scattering on the surface.}

The high-frequency operation of acoustic devices is important for RF application as well as for opto-acoustic technology, where the high-frequency acoustic modulation of semiconductor optical properties enables the control of optical signals in a photonic chip \cite{macquarrie2015coherent,chen2019engineering, tian2020hybrid,liu2020monolithic}. The possible excitation frequency of the transducer is roughly defined by
\begin{eqnarray} 
	f_n \sim n\times \left( \frac{v_\text{p}}{2t_\text{p}} \right) ;(n=1, 3, 5\cdots),
	\label{f_n}
\end{eqnarray}
where $v_\text{p}$ and $t_\text{p}$ respectively denote longitudinal acoustic velocity and thickness of the piezoelectric layer. Since an actual transducer has a top electrode, the thickness-mode frequency shifts from that obtained with Eq. (\ref{f_n}) due to the mass loading effect.
We fabricated two other versions of our HBAR having 400- and 200-nm piezoelectric layer thicknesses by post-growth etching. 
The AlN layer was etched with inductively coupled plasma (ICP) using Ar/Cl$_2$/BCl$_3$ (5:2:13) gas chemistry. After the ICP etching, the top electrode of Ti/Al (5/100 nm) was deposited by electron beam evaporation.
\textcolor{black}{We observed the surface morphology of unetched and etched AlN film by atomic force microscopy. Root mean square roughness of films was 0.450 (800-nm AlN), 0.656 (400-nm AlN), and 0.961 nm (200-nm AlN) (see Supplementary material). The longer the etching time, the larger the surface roughness.}

Figure 3(a) shows the reflection spectrum when the AlN layer was etched to the thickness of 400 nm. The center frequency of the reflection-spectrum envelope was around 8 GHz, and this envelope corresponds to the fundamental thickness mode of the transducer. 
When the thickness of the AlN layer was reduced to 200 nm, the center frequency shifted to 10 GHz, as shown in Fig. 3(b).
The high figure of merit of $f\times\text{Q} = 1.3\times 10^{13}\ \textrm{Hz}$ was achieved even at the high frequency of 10 GHz due to the high quality of the epitaxial AlN layer. 
We observed the envelope of the third thickness-mode in the thickness-reduced HBARs. 
The inset of Fig. 3(b) shows an enlarged spectrum of the HBAR mode near 26.5 GHz. Since the possible excitation frequency was limited by the VNA used in our measurements, we will find the higher-order thickness mode in the extremely high-frequency region by using appropriate measurement systems.
The $f\times\text{Q}$ products of the 200- and 400-nm HBARs are respectively shown in Figs. 3(c) and 3(d).
\textcolor{black}{Note that the Q-factors of the third thickness-mode were also measured by taking the average of 2000 measurements.}
\textcolor{black}{The $f\times\text{Q}$ products of the 200-nm HBAR are lower than the other two HBARs. This might be because the larger surface roughness and thick top metal electrode degrade the Q-factor \cite{hashimoto2009rf}.}
\textcolor{black}{HBARs with as-grown AlN thin film should show superior properties compared with HBARs with etched AlN film due to improved surface flatness.}

\begin{figure}[th]
	\centering
	\includegraphics[width=9cm]{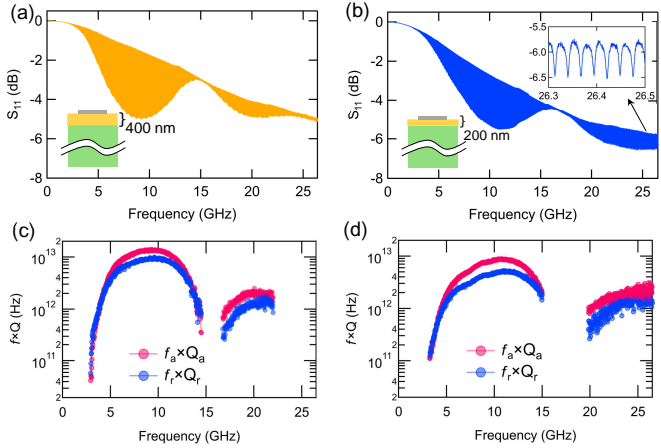}
	\caption{High-frequency operation. As-grown thickness of 800 nm was reduced to 400 (a) and 200 nm (b) by ICP etching. Inset of (b) shows enlarged view, where HBAR modes were clearly observed.  (c)$-$(d) Frequency dependence of $f\times\text{Q}$ product, where thickness of AlN was 400 (c) and 200 nm (d).}
	\label{f3}
\end{figure} 

The FSR of HBARs with multiple phonon modes is defined by
\begin{eqnarray} 
	\text{FSR}\left(m\right) = f_{m+1}-f_m.
\end{eqnarray}
The quasi-periodic nature of the FSR spectrum has been explored  \cite{gokhale2020engineering,kongbrailatpam2020effects, sandeep2017resonant, zhang2003resonant, Sorokin2021Toward, chen2006characterization}. When the acoustic impedance between the transducers and acoustic cavities nearly matches, the FSR spectrum shows small sinusoidal amplitude variation. If the entire HBAR has the same acoustic impedance, as illustrated at the bottom of Fig 4(a), the FSR exhibits a flat spectrum, shown as the green line in Fig. 4(a).
In contrast, the FSR spectra exhibit periodic delta functions with large modulation amplitude when the impedance mismatch is large (see Supplementary material). 
Therefore, we can experimentally verify that an HBAR has good impedance matching by measuring the variation in FSR. Figure 4(a) also shows the experimental FSR spectra with respect to the frequency in our HBAR with an AlN thickness of 800 nm and the active area of 10000 $\mu$m$^2$.
The gray line indicates the experimental FSR and blue line is the smoothed experimental FSR. 
\textcolor{black}{It should be noted that the conductance of the electrode does not affect the FSR spectrum\cite{Lakin2001improved}.}

\begin{figure}[bth]
	\centering
	\includegraphics[width=9cm]{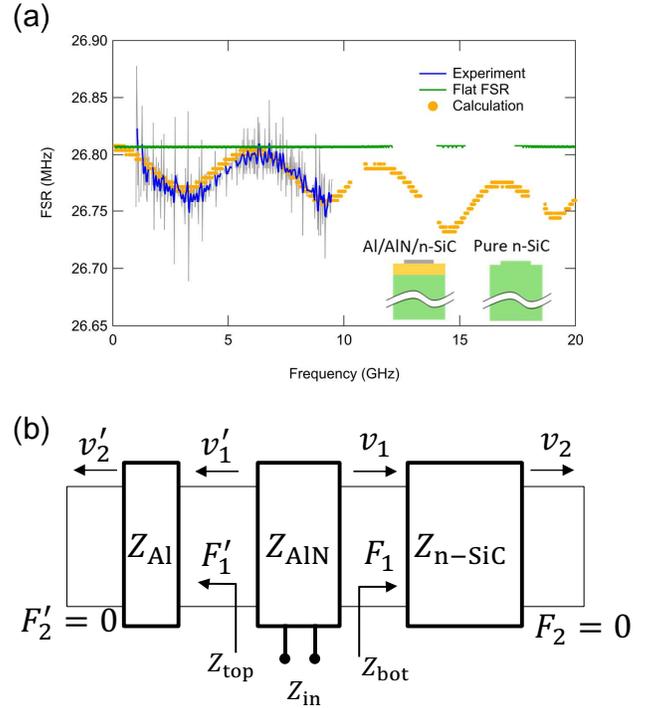}
	\caption{Variation in FSR. (a) FSR spectra. Gray line corresponds to experimental results, where finite noise was superposed by environmental noise and transverse modes. Blue line represents smoothed experimental results, where Gaussian smoothing with the number of 2 smoothing operations were applied.  Orange circles are deduced from numerical calculations with the Mason model.(b) Matrix model.}
	\label{f4}
\end{figure}  
\begin{table*}[ht]
	\caption{Comparison of calculated $C_\text{FSR}$s}
	\centering
		\begin{ruledtabular}
			\begin{tabular}{ccccccc}
				HBAR  & Material stack & Fundamental mode (GHz) & $\sigma_\text{FSR}$ (kHz)  &  $\mu_\text{FSR}$ (MHz) &   $C_\text{FSR}$ \\
				\hline 
				Ref [10] & M$^*$/BSTO/M/Sapphire & 0.6 & 33.6 & 10.8  & 3.1$\times 10^{-3}$\\
				Ref [22]  & M/ZnO/M/SiO$_2$ & 1.3 & 5  & 1.94  & 2.6$\times 10^{-3}$ \\
				Ref [18]   & M/AlScN/M/Diamond & 2 & 45.6  & 17.2  & 2.6$\times 10^{-3}$ \\
				Ref [21]   & M/GaN/AlN/M/SiC & 3.1 & 15.5  & 18.9  & 8.2$\times 10^{-4}$ \\
				Ours  & M/AlN(800 nm)/n-SiC & 5.9 & 18.7 & 26.8  & 7.0$\times 10^{-4}$ \\
				Ours  & M/AlN(400 nm)/n-SiC & 8 & 15.2 & 26.9  & 5.7$\times 10^{-4}$ \\
				Ours  & M/AlN(200 nm)/n-SiC & 10 & 13.0  & 27.0  & 4.8$\times 10^{-4}$ \\
				& *M: Metal & & & & \\
			\end{tabular}
			\label{t1}
		\end{ruledtabular}
	\end{table*}
For modeling our HBARs, we used the one-dimensional Mason model \cite{zhang2003resonant}, with which the electrical input impedance is calculated.
The schematic matrix model of the three-layer composite resonator is shown in Fig. 4(b), where we omitted the adhesion layer of Ti for simplicity. The acoustic impedance at the left side of the piezoelectric layer is expressed as
\begin{eqnarray} 
	Z_{\text{top}} = \frac{F_1^{\prime}}{v_1^{\prime}} = jZ_{\text{Al}}\tan(k_{\text{Al}}d_{\text{Al}}),
	\label{Z_Al}
\end{eqnarray}
where $F$ and $v$ respectively represent force and velocity, and $k_i$ and $d_i$ respectively denote the wavenumber and thickness. It should be noted that forces at the interface between the top (bottom) layers and air, $F_2^{\prime}$ ($F_2$), is zero, which corresponds to an electric short in the circuit.
As the schematic matrix model is symmetrical with respect to the piezoelectric layer, the $Z_{\text{bot}}$ has the same acoustic impedance formula as Eq. (\ref{Z_Al}), but the material property is that of the n-SiC. As a result, the electric input impedance of the three-layer composite resonator is expressed as \cite{zhang2003resonant}
\begin{eqnarray} 
	Z_\text{in}=\frac{1}{j\omega C_0}\left[1-\frac{k_t^2}{\gamma}
	\frac{(z_{\text{top}}+z_{\text{bot}})\sin{\gamma}+j2(1-\cos{\gamma}) }
	{(z_{\text{top}}+z_{\text{bot}})\cos{\gamma}+j(1+z_{\text{top}} z_{\text{bot}})\sin{\gamma}}\right],
	\label{Z_in}
\end{eqnarray}
where $\gamma = k_{\text{AlN}}d_{\text{AlN}}$ and $C_0 = \varepsilon_{33} S/t$. $\varepsilon_{33}$ and $t$ are respectively the permittivity and thickness of the piezoelectric layer, and $k_t^2$ is the intrinsic electromechanical coupling coefficient of the AlN and assumed as 6.5 \% in the numerical calculation.
The $z_{\text{top}}=Z_{\text{top}}/Z_\text{AlN}$ and $z_{\text{bot}}=Z_{\text{bot}}/Z_\text{AlN}$ are respectively the normalized acoustic impedance of the top electrode and acoustic cavity by the acoustic impedance of the AlN layer. 

The numerically calculated FSR of the 800-nm AlN HBAR is shown as the orange circles in Fig. 4(a) showing the small amplitude of sinusoidal variation due to the small impedance mismatch between the transducer and acoustic cavity. 
Although the experimental results indicate fluctuations due to the finite environmental noise and the existence of transverse modes, the smoothed experimental results are in good agreement with the calculated results.
The experimental results are shown below 10 GHz because the signal above 10 GHz in our 800-nm HBAR was so weak that we could not identify the resonance frequency.

Finally, we evaluated the FSR variation to confirm that our HBAR has an impedance matched structure.
To compare the FSR variation with various other HBARs, we introduced the normalized FSR variation factor defined by
\begin{eqnarray} 
	C_\text{FSR} = \frac{\sigma_\text{FSR}}{\mu_\text{FSR}},
\end{eqnarray}
where $\sigma_\text{FSR}$ is the standard deviation of the FSR, and $\mu_\text{FSR}$ is the average of the FSR.
Normalization by $\mu_\text{FSR}$ is necessary since an HBAR with a thicker substrate has a smaller $\sigma_\text{FSR}$ when the film-thickness ratio is the same.
The $C_\text{FSR}$ of our 800-nm AlN HBAR was $8.7\times 10^{-4}$, which is deduced from the unsmoothed experimental results.
We compared the calculated $C_\text{FSR}$ among the previous HBARs, as shown in Table \ref{t1}, where the frequency range is below 20 GHz.
The FSR variation was calculated using the Mason model on the basis of the material information in the previous studies.  
The $C_\text{FSR}$ of our HBARs with 800-, 400-, and 200-nm AlN layers were 7.0$\times 10^{-4}$,  5.7$\times 10^{-4}$, and  4.8$\times 10^{-4}$, respectively.
The calculated $C_\text{FSR}$ is plotted against the experimentally obtained center frequency of the fundamental thickness mode in Fig. 5, where the blue circles and red stars denote the previous and our HBARs, respectively.
Our HBAR potentially has the lowest  $C_\text{FSR}$. 
\begin{figure}[th]
	\centering
	\includegraphics[width=7cm]{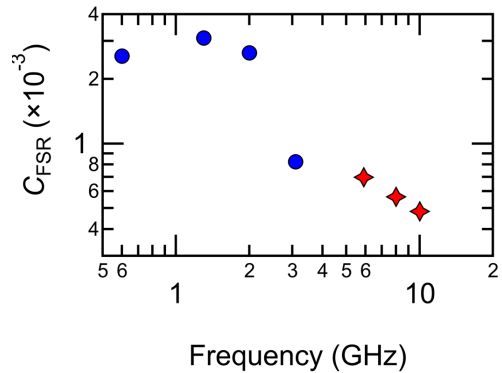}
	\caption{(a) Comparison of calculated $C_\text{FSR}$s plotted from Table 1. Horizontal axis corresponds to center frequency of fundamental thickness mode obtained from measurement. Blue circles and red stars correspond to previously reported and our HBARs, respectively.}
	\label{f5}
\end{figure} 

In conclusion, we fabricated a nearly impedance-matched HBAR in which the piezoelectric AlN layer was epitaxially grown on a conductive n-SiC substrate. In contrast to conventional HBARs using a bottom metal electrode, the n-SiC substrate of our HBAR functions both as a floating bottom electrode and acoustic cavity. 
We confirmed that the experimental spectra of FSR is well described using the one-dimensional Mason model. 
The experimental $C_\text{FSR}$ ($8.7\times 10^{-4}$) is comparable to the best value in other HBARs.
The small impedance mismatch was verified by the small variation in the FSR.
The calculated $C_\text{FSR}$ of our HBAR was 4.8$\sim$7.0 $\times 10^{-4}$, which is the lowest in HBARs. 
The high-frequency operation in the K-band ($<$ 26.5 GHz) was also demonstrated by reducing the thickness of the AlN layer. The high figure of merit of $f\times\text{Q} = 1.3\times 10^{13}\ \textrm{Hz}$ at 10 GHz was achieved due to the high quality of the epitaxial AlN layer.
\textcolor{black}{AlN is a wide bandgap semiconductor used in ultraviolet photonic devices and RF electronic devices \cite{taniyasu2006aluminium,hiroki2022high}. An epitaxially grown AlN is especially promising for integrating acoustic devices with these optical and RF electronic devices on the same chip \cite{Zhao2022xband}.}

Using conductive substrates as electrodes will lead to perfectly impedance-matched HBARs, in which the same materials are used as the piezoelectric layer and doped-semiconductor electrode such as n-GaN/GaN/n-GaN. This HBAR exhibits flat FSR spectra and has an optically transparent structure, which is advantageous for
the hybrid coupling between high-frequency phonon modes and optical systems. 
These results will pave the way to the fundamental research of high-frequency acoustic systems hybridized with electric, optical, and magnetic systems.

\section*{Supplementary material}
See supplementary material for more information.

\begin{acknowledgments}
	The authors thank Paulo V. Santos, Kazuhide Kumakura, and Keiko Takase for fruitful discussions.
\end{acknowledgments}

\section*{Data Availability}
The data that support the findings of this study are available from the corresponding author upon reasonable request.

%

\end{document}